\documentclass[floatfix, noshowpacs, preprintnumbers, twocolumn, amsmath, amssymb, aps, superscriptaddress, prb]{revtex4}

\pdfoutput=1

\usepackage{graphicx, xcolor}
\usepackage[caption=false]{subfig}
\usepackage{soul}

\newcommand{\ket}[1]{\left| #1 \right>}

\begin{document}

\title{Experimental quantum stochastic walks simulating associative memory\\
 of Hopfield neural networks}

\author{Hao Tang}
\affiliation{School of Physics and Astronomy, Shanghai Jiao Tong University, Shanghai 200240, China}
\affiliation{Institute of Natural Sciences, Shanghai Jiao Tong University, Shanghai 200240, China}
\affiliation{Synergetic Innovation Center of Quantum Information and Quantum Physics, University of Science and Technology of China, Hefei, Anhui 230026, China}

\author{Zhen Feng}
\affiliation{School of Physics and Astronomy, Shanghai Jiao Tong University, Shanghai 200240, China}
\affiliation{Synergetic Innovation Center of Quantum Information and Quantum Physics, University of Science and Technology of China, Hefei, Anhui 230026, China}

\author{Ying-Han Wang}
\affiliation{School of Physics and Astronomy, Shanghai Jiao Tong University, Shanghai 200240, China}
\affiliation{Zhiyuan Innovation Research Center, Shanghai Jiao Tong University, Shanghai 200240, China}

\author{Peng-Cheng Lai}
\affiliation{School of Physics and Astronomy, Shanghai Jiao Tong University, Shanghai 200240, China}
\affiliation{Zhiyuan Innovation Research Center, Shanghai Jiao Tong University, Shanghai 200240, China}

\author{Chao-Yue Wang}
\affiliation{School of Physics and Astronomy, Shanghai Jiao Tong University, Shanghai 200240, China}
\affiliation{Zhiyuan Innovation Research Center, Shanghai Jiao Tong University, Shanghai 200240, China}

\author{Zhuo-Yang Ye}
\affiliation{School of Physics and Astronomy, Shanghai Jiao Tong University, Shanghai 200240, China}
\affiliation{Zhiyuan Innovation Research Center, Shanghai Jiao Tong University, Shanghai 200240, China}

\author{Cheng-Kai Wang}
\affiliation{School of Physics and Astronomy, Shanghai Jiao Tong University, Shanghai 200240, China}
\affiliation{Zhiyuan Innovation Research Center, Shanghai Jiao Tong University, Shanghai 200240, China}

\author{Zi-Yu Shi}
\affiliation{School of Physics and Astronomy, Shanghai Jiao Tong University, Shanghai 200240, China}
\affiliation{Synergetic Innovation Center of Quantum Information and Quantum Physics, University of Science and Technology of China, Hefei, Anhui 230026, China}

\author{Tian-Yu Wang}
\affiliation{School of Physics and Astronomy, Shanghai Jiao Tong University, Shanghai 200240, China}

\author{Yuan Chen}
\affiliation{School of Physics and Astronomy, Shanghai Jiao Tong University, Shanghai 200240, China}
\affiliation{Synergetic Innovation Center of Quantum Information and Quantum Physics, University of Science and Technology of China, Hefei, Anhui 230026, China}
\affiliation{Institute for Quantum Science and Engineering and Department of Physics, Southern University of Science and Technology, Shenzhen 518055, China}

\author{Jun Gao}
\affiliation{School of Physics and Astronomy, Shanghai Jiao Tong University, Shanghai 200240, China}
\affiliation{Synergetic Innovation Center of Quantum Information and Quantum Physics, University of Science and Technology of China, Hefei, Anhui 230026, China}
\affiliation{Institute for Quantum Science and Engineering and Department of Physics, Southern University of Science and Technology, Shenzhen 518055, China}

\author{Xian-Min Jin}
\email{xianmin.jin@sjtu.edu.cn} 
\affiliation{School of Physics and Astronomy, Shanghai Jiao Tong University, Shanghai 200240, China}
\affiliation{Institute of Natural Sciences, Shanghai Jiao Tong University, Shanghai 200240, China}
\affiliation{Synergetic Innovation Center of Quantum Information and Quantum Physics, University of Science and Technology of China, Hefei, Anhui 230026, China}

\email{xianmin.jin@sjtu.edu.cn} %% email address is required

\begin{abstract}
With the increasing crossover between quantum information and machine learning, quantum simulation of neural networks has drawn unprecedentedly strong attention, especially for the simulation of associative memory in Hopfield neural networks due to their wide applications and relatively simple structures that allow for easier mapping to the quantum regime. Quantum stochastic walk, a strikingly powerful tool to analyze quantum dynamics, has been recently proposed to simulate the firing pattern and associative memory with a dependence on Hamming Distance. We successfully map the theoretical scheme into a three-dimensional photonic quantum chip and realize quantum stochastic walk evolution through well-controlled detunings of the propagation constant. We demonstrate a good match rate of the associative memory between the experimental quantum scheme and the expected result for Hopfield neural networks. The ability of quantum simulation for an important feature of a neural network, combined with the scalability of our approach through low-loss integrated chip and straightforward Hamiltonian engineering, provides a primary but steady step towards photonic artificial intelligence devices for optimization and computation tasks of greatly improved efficiencies.
\end{abstract}

\maketitle
\vskip -3.5mm
\section{INTRODUCTION}
\vskip -3.5mm
The crossover between quantum information and machine learning has been an emerging field\cite{Biamonte2017} to generate both the quantum-enhanced machine learning tasks\cite{Lloyd2014, Rebentrost2014, Cai2015, Li2015, Spagnolo2017} and the machine-learning-assisted quantum algorithms\cite{Carrasquilla2017, vanNieuwenburg2017, Lu2017,Gao2017}. The former utilizes the quantum superposition to speed up the machine learning performance, which has been recently studied in tasks such as principal component analysis (PCA) \cite{Lloyd2014} and support vector machines (SVM) \cite{Rebentrost2014,Cai2015,Li2015}. The latter applies machine learning techniques to quantum problems ranging from the phase transition in condensed matter physics \cite{Carrasquilla2017, vanNieuwenburg2017} to the quantum state classification\cite{Lu2017,Gao2017}, in order to improve the data processing efficiency for complex quantum systems. 

Among the various machine learning techniques, artificial neural networks have drawn lots of attention due to their versatile utilities for deep learning \cite{LeCun2015,Silver2016,Minh2015} and neuromorphic computing\cite{Merolla2014,Shen2017}. Similar to those in human brain, artificial neural networks consist of neuron structures, and include mechanisms on attractor dynamics, firing pattern, training rules, etc. A very neat and popular type of neural networks is Hopfield network\cite{Hopfield1982,Hopfield1986}, a single-layered and recurrent neural network with undirected connections between neurons. The signal transmission between neurons is always directed, from the synapse in one neuron to the dentrites in the other neuron, but the bidirectional connections between two neurons can still be realized by mutually receiving signals from the other's synapse, as illustrated in Fig.1.a. With flexible connections among neurons, Hopfield neural network presents a prominent feature, the associative memory, which directly drives the network to form the firing pattern at the energy minima closest to the initial input pattern (Fig.1.b). Such a content-addressable memory system can be widely applied to optimization, image processing\cite{Tatem2001,Zhu1997}, and even identifying genetic segments of RNA\cite{Rebentrost2018}, where the distance between patterns can be quantified by Hamming Distance and the energy minima become the optimized task solution. Therefore, even though Hopfield network is not as fancy as many trendy deep learning networks, quantum simulations for the former are still appealing, because it shows the important associative memory and its relatively simple structure allows for an easier mapping into the quantum regime that can serve as building blocks for more advanced quantum structures\cite{Rebentrost2018}.

\begin{figure*}[ht!]
\includegraphics[width=1.0\textwidth]{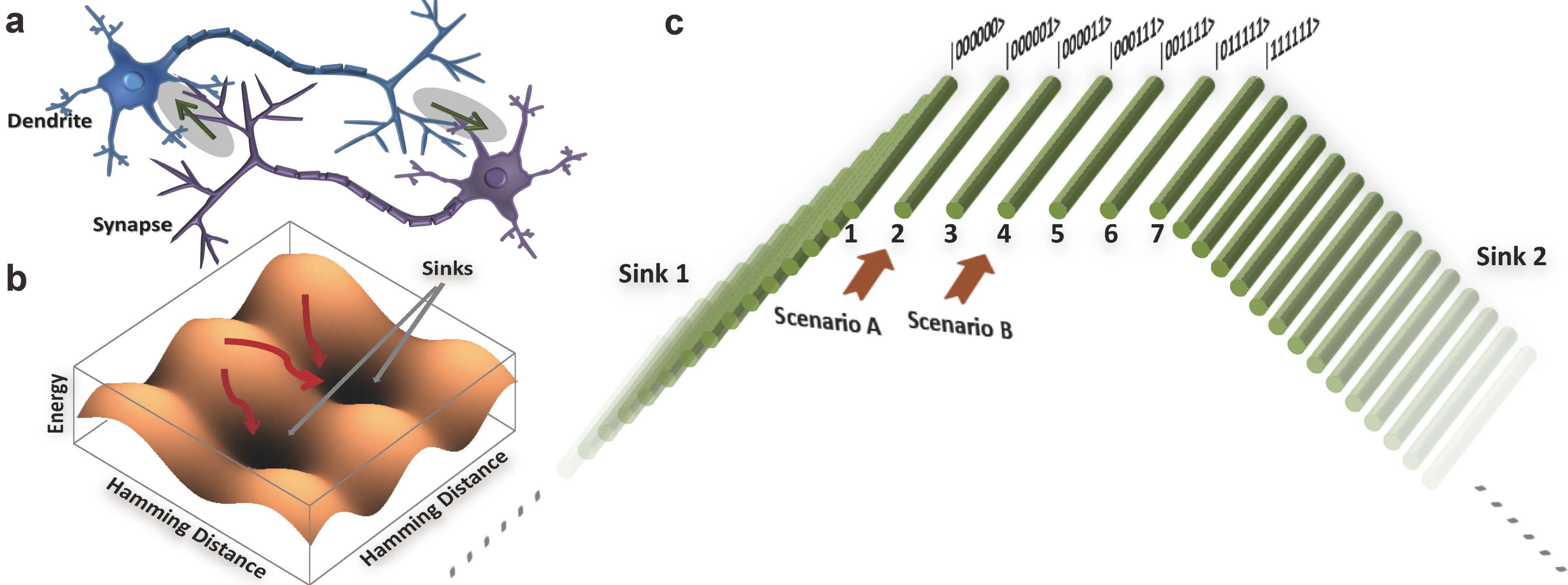}
\caption{\textbf{The Hopfield network and the mapping to integrated photonics.} ({\bf a}) Schematic diagram of bidirectional signal transmission between two neurons. The green arrows show the transmission from the synapse in one neuron to the dentrite in the other neuron. ({\bf b}) Schematic diagram of the Hopfield network that memorizes the energy minima that are closest to the initial firing pattern in terms of Hamming Distance. ({\bf c}) Schematic diagram of a seven-state network in the photonic waveguide array. The Waveguide 1-7 represent the seven states, and the long array of waveguides on the side represent the sink for State1 and State 7, respectively. Two scenarios of initial patterns are considered, with the initial firing state being State 2 only in Scenario A and being State 4 only in Scenario B.}
\label{fig:QFTConcept}
\end{figure*}

Quantum neural networks have hence been proposed and were hoped to fully comply with both neural network mechanisms and quantum theory. However, this is confronted with a severe dilemma that the non-linear activation functions inherently contradict the linear evolution in quantum theory\cite{Schuld2014b}. Some recent theoretical proposals stick to complete neural networks using universal quantum gates\cite{Rebentrost2018,Cao2017,Wan2017}, but their experimental realization could be too challenging at this stage, even for one single neuron, let alone a considerably large number of neurons that are required to make the neural network really work. On the other hand, some alternative approaches that simulate partial neural features, e.g., the quantum associative memory\cite{Ventura2000,Trugenberger2001, Diamantini2015}, have been more extensively studied in theory. A model of quantum stochastic walks, a mixture of quantum and classical walk\cite{Whitfield2010}, has been raised to simulate the firing pattern and associative memory\cite{Schuld2014}. Although it is worth emphasizing that such simulation does not realize the real quantum neurons, yet the mapping between neural networks and quantum stochastic walks has already been of great interest: It is a strong crossover between a representative neural feature and a very versatile quantum approach, thus bridging two separated fields together and greatly broadening the applications for each field. So far, this has been limited in theories only, but with the experimental advances of quantum walks\cite{Du2003,Schmitz2009,Perets2008,Schreiber2012,Jeong2013, Tang2018} and quantum stochastic walks\cite{Biggerstaff2016, Caruso2016}, its experimental demonstration becomes much more appealing and promising.
\vskip -3.5mm
\section{RESULTS}
\vskip -3.5mm
Therefore, in this letter, we experimentally present the quantum stochastic walk in a three-dimensional (3D) photonic quantum chip and use it to demonstrate the feature of associative memory. In our setting, as shown in Fig.1.c, we propose a seven-state network represented by an array of seven waveguides that sequentially correspond to the following states: $\ket{000000}$, $\ket{000001}$, $\ket{000011}$, $\ket{000111}$, $\ket{001111}$,  $\ket{011111}$ and $\ket{111111}$. Each state differs from its neighboring state by a Hamming Distance of one, that is, only one different symbol in the same position of the two binary strings. In a mapping to waveguide arrays, the Hamming Distance is equivalent to the count of waveguide spacings. For example, Waveguide 1 which represents State 1 ($\ket{000000}$) has a Hamming Distance of six (or six waveguide spacings) from Waveguide 7 representing State 7 ($\ket{111111}$). The mutually evanescent coupling between neighbouring photonic waveguides simulates the undirected connections in a Hopfield network. Now we make State 1 and State 7 as two dynamic basins, or the `sink states', by connecting to Waveguide 1 and Waveguide 7 each a long array of 50 auxiliary waveguides (See APPENDIX for fabrication details). The spacing for adjacent auxiliary waveguides is much smaller than that for the seven state waveguides, yielding a much stronger coupling coefficient for the former, so photons would pass other states and evolve to the two sink states. 

As suggested by Schuld {\it et al } \cite{Schuld2014}, the network evolution can be simulated by a quantum stochastic walk:
$$\frac{d\rho}{dt}=-(1-\omega)i[H,\rho]+\omega\sum_{ij}(L_{ij} \rho L_{ij}^\dagger- \frac{1}{2}\{L_{ij}^\dagger L_{ij}, \rho\})\eqno{(1)}$$  
where the first part right to the equation (containing $i[H,\rho]$) is the coherent and symmetric Hamiltonian term, and the second part (containing $L_{ij}$ and $L_{ij}^\dagger$) is classical Lindblad term that causes diagonal and non-coherent environment noises in the open quantum system. In this model\cite{Schuld2014}, the connections between sites are undirected, while adding an extra directed site in the Lindblad term creates a sink of an energy minimum, and the noises from the Lindblad term assist and enhance the direct jump to those sinks.

\begin{figure}[ht!]
\includegraphics[width=0.49\textwidth]{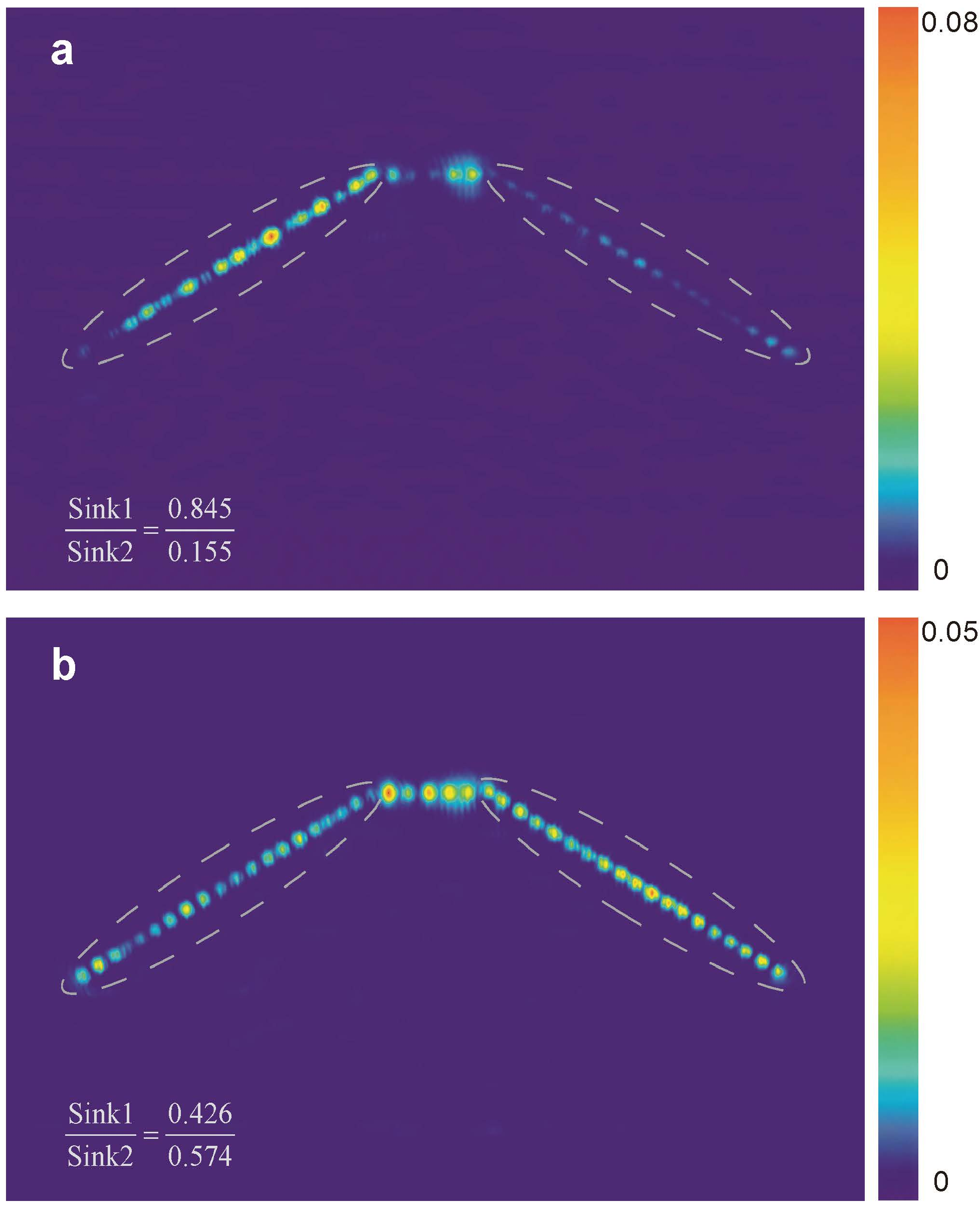}
\caption{\textbf{Experimental evolution patterns.} An experimental evolution pattern  ({\bf a}) for Scenario A, with initial photon injection in State 2, ({\bf b}) for Scenario B with initial photon injection in State 4. Both are from one sample with a $\Delta \beta_A$ of 0.4~$\rm mm^{-1}$. For each scenario, the dashed ellipses in the left and right wing correspond to Sink 1 and Sink 2, respectively, and the inserted data shows the ratio of the light intensity in Sink 1 and that in Sink 2. 
}
\label{fig:strutturaChip}
\end{figure} 

A photonic model that realizes such quantum stochastic walks has been proposed\cite{Caruso2016}. It introduces quantitatively controllable diagonal decoherent terms into a photonic quantum system, and was demonstrated for a `maze' escaping problem\cite{Caruso2016}. This method, which can actually be quite useful for simulating many problems in open quantum systems, had never been used in experiments after its invention. We manage to realize the experimental implementation once again, and apply it to an entirely new field, the simulation of a machine learning feature.

In the photonic model, the diagonal and non-coherent noise can be created effectively by controlling $\Delta \beta$, which is the detuning of propagation constant in the diagonal term of the Hamiltonian matrix, while the long array of auxiliary waveguides can be employed to create the sinks. In light of this $\Delta \beta$ model, we randomly pick $\Delta \beta$ from the uniform distribution of a certain amplitude, $\Delta \beta_A$. We fabricate the waveguide structure using femtosecond laser direct writing and realize the $\Delta \beta$ by adjusting the writing speed (see APPENDIX for more details). We set a few $\Delta \beta_A$ in different samples of the same array structure to vary the contribution of classical walk, since it has been qualitatively demonstrated that\cite{Caruso2016}, a higher detuning amplitude of the propagation constant corresponds to a larger portion of the Lindblad terms in the quantum stochastic walk.

\begin{figure*}[ht!]
\includegraphics[width=1\textwidth]{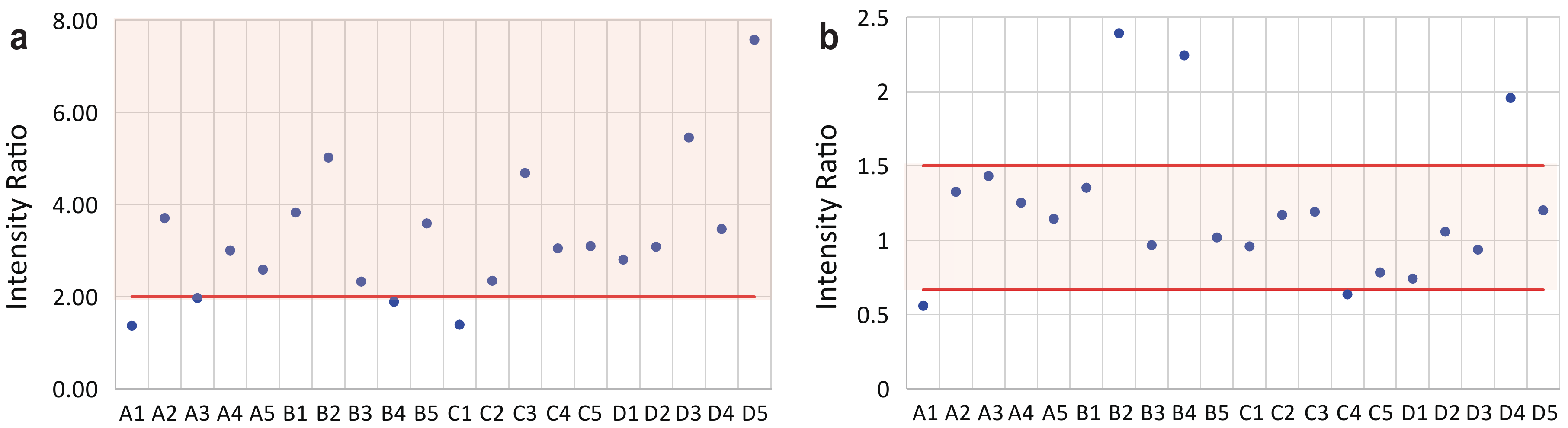}
\caption{\textbf{ The match rate for associative memory.} The experimentally characterized light intensity ratio of Sink 1 and Sink 2 for Scenario A in ({\bf a}) and Scenario B in ({\bf b}) for all twenty samples. The shadowed areas show the criteria of the correct simulation of associative memory for each scenario. The match rate for associative memory can be counted as the percentage of dots in the shadowed area over the total number of dots in each each scenario.}
\label{fig:apparato}
\end{figure*}

\begin{figure*}[t!]
\includegraphics[width=0.65\textwidth]{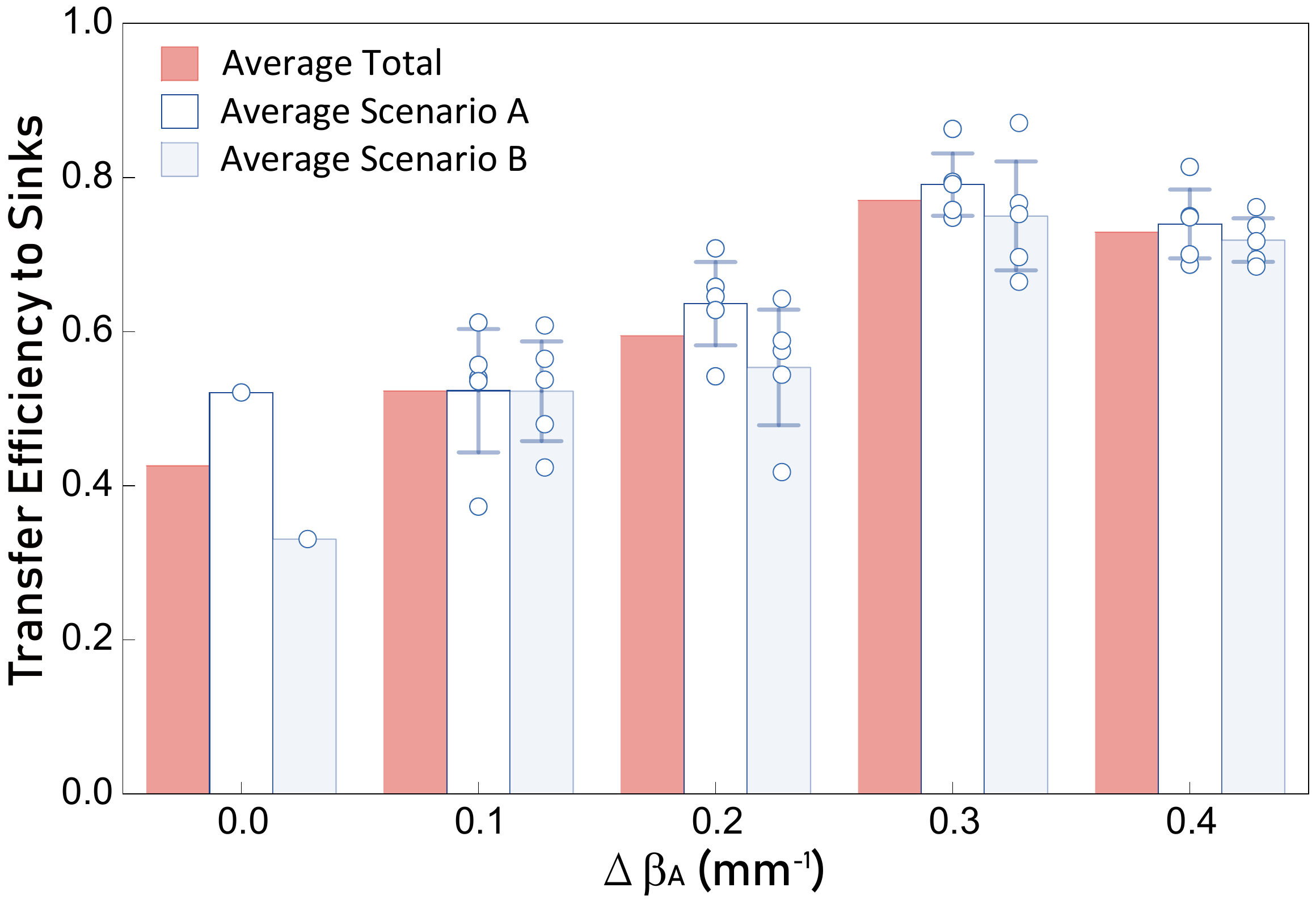}
\caption{\textbf{The transfer efficiencies.} The experimental transfer efficiencies to the sink states for samples with different $\Delta \beta_A$ in ({\bf a}) Scenario A and ({\bf b}) Scenario B. The small circles represent the individual result for each sample. The columns show the average for each group of the same $\Delta \beta_A$, and the attached error bars are calculated by the standard variance of the five individual results within each group. Average Total stands for the average value for all results from both Scenario A and Scenario B in the same group of $\Delta \beta_A$.}
\label{fig:Results4}
\end{figure*}

We now consider two representative scenarios of initial patterns, where the initial firing state is State 2 only in Scenario A, and State 4 only in Scenario B. The two scenarios differ in their Hamming Distance between the initial state and the sink states, that is, State 2 is closer to State 1 than to State 7, whereas State 4 has an equal Hamming Distance to State 1 and State 7. We inject the 780 nm vertically polarized laser beam into Waveguide 2 and Waveguide 4 to simulate Scenario A and Scenario B, respectively. We measure the evolution patterns using a CCD camera, with an example shown in Fig.2 and calculate the ratio of light intensity at Sink 1 and Sink 2. 

It has been theoretically proven by both the Hopfield neural network model and the quantum stochastic walk model\cite{Schuld2014}, that the walk always fully evolves to the sink state closest to the initial state in terms of the Hamming Distance, and if there are two sink states of an equal Hamming Distance to the initial state, the walk will end up with equal probabilities at the two sink states. Most firing patterns can be classified into either of these two representative scenarios. In experiment, the probability distribution is reflected by the light intensity, and the retrieved results agree with theory well. To quantitatively evaluate the performance, we propose that, if the light intensity in sink State 1 is more than two times of that in sink State 7 for Scenario A, within certain experimental tolerance, the walk still has a clear preference for sink State 1 indicating the simulation of associative memory for Scenario A is correct; whereas for Scenario B, we regard the simulation is correct if the light intensity ratio stays between 40$\%$:60$\%$ and 60$\%$:40$\%$ that suggests a roughly balanced distribution. The inserted values in Fig.2.a and b both show a correct simulation result. 
 
We investigate the match rate of associative memory, as shown in Fig.3. For both Scenario A (Fig.3.a) and Scenario B (Fig.3.b), we prepare 20 samples and analyze their light intensity ratio from the measured patterns. Among the 20 samples we've prepared, the configurations of the waveguide structure are identical but $\Delta \beta$ values are different: there are four groups of $\Delta \beta_A$ as 0.1, 0.2, 0.3 and 0.4  $\rm mm^{-1}$ for Group A-D respectively, and each group includes five samples which have various random $\Delta \beta$ values under the same $\Delta \beta_A$. The dots in the shadowed range are those that show the correct simulation result, while the dots outside the range do not match our tight criteria, so we can count that Scenario A and B have a match rate of the associative memory as 80\% and 75\%, respectively. The quantum stochastic walk on the photonic waveguide system has thus demonstrated positive results for simulating the associative memory effect that's consistent with the theory \cite{Schuld2014}. 

From Fig.3 the match rate does not show clear differences among groups of different delta beta amplitudes, which suggests a robust simulation of associative memory insensitive to the parameter of the quantum model. Additionally, the change of $\Delta \beta_A$ does have some other effects. As predicted in the theoretical paper\cite{Schuld2014}, the proportion of classical term in the quantum stochastic walk influences the transfer efficiency at the same evolution time. Correspondingly, we measure the portion of light intensity in the sink State 1 and 7 over the total light intensity as the transfer efficiency to the sink states, and plot the transfer efficiency for samples of different $\Delta \beta_A$ in Fig 4. For both Scenario A and B, the increase of $\Delta \beta_A$ clearly enhances the transport efficiency comparing to the pure quantum walk where $\Delta \beta_A$ equals zero. A slight drop of the transport efficiency when $\Delta \beta_A$ further increases to 0.4 $\rm mm^{-1}$ is consistent with the theoretical prediction that an optimal transfer efficiency exists at a certain classical amount of the walk. From the feature of the enhanced transfer efficiencies over pure quantum walks, we have verified the positive role of classical terms in speeding up the formation of the associative memory in quantum stochastic walks. 

\vskip -3.5mm
\section{CONCLUSION}
\vskip -3.5mm
Through this work, we have presented the first attempt to experimentally simulate the associative memory of Hopfield networks using quantum stochastic walks on photonic waveguide arrays. The simulation does not realize full features of neurons, but focuses on a most important feature of Hopfield network, the associative memory. Just like Hopfield network `memorizes' the dynamic basin that's close to the initial pattern in terms of the Hamming Distance, we use the quantum stochastic walk of photons to `memorize' the correct sinks dependent on the waveguide spacing. We have a faithful implementation of the theoretical proposal by demonstrating the theoretically proposed two different and very representative types of firing patterns that most scenarios can be classified into. Both the theoretical proposal and our experimental results suggest a modest speedup through the approach of quantum stochastic walk. Even though the speedup is not extremely high, we manage to set the first real step forward towards its experimental demonstration and this could serve as a building block for future quantum devices.

It is further inspiring to see the scalability of our approach for this work. In fact, the theoretical proposal itself has scaling advantages. It uses a very plain mapping into the quantum stochastic walks, which, even for very large evolution spaces, can be straightforwardly implemented in photonic systems. One may concern if a larger scale in experiments would cause losses to destroy the originally modest quantum speedup advantages, noticing that error correction is critical for universal quantum computing, and some approach of using phase-shifters \cite{Carolan2015, Harris2017, Sparrow2018} to construct a designed unitary matrix has only limited scales reported, because of stringent fabrication requirements and a polynomially increasing number of phase-shifters when the scale increases. On the other hand, our approach is to flexibly construct the Hamiltonian matrix that contains the coupling information through defining the waveguide structure directly. The scalability for such a scheme could benefit from the larger-scale and lower-loss evolution system that can be straightforwardly laid out on integrated photonic chips, and the flexible and precise setting of coupling coefficients through curvature, space control, etc. These allow for the potential feasibililty for wider associative memories in more complex firing patterns, and a scalable solution to another task, experimental quantum fast hitting, has also been recently demonstrated \cite{Tang2018b}. In future, more tasks can be mappped into the Hamiltonian matrix on a photonic quantum chip using such an analog quantum computing approach. Our attempt as an elementary example brings up great many possibilities for scalable analog quantum computing applications worthy of further explorations. \\

\vskip -3.5mm
\vskip -3.5mm
\section*{ACKNOWLEDGMENTS}
\vskip -3.5mm
The authors thank Roberto Osellame and Jian-Wei Pan for helpful discussions. This work was supported by National Key R\&D Program of China (2017YFA0303700); National Natural Science Foundation of China (NSFC) (61734005, 11761141014, 11690033); Science and Technology Commission of Shanghai Municipality (STCSM) (15QA1402200, 16JC1400405, 17JC1400403); Shanghai Municipal Education Commission (SMEC)(16SG09, 2017-01-07-00-02-E00049); X.-M.J. acknowledges support from the National Young 1000 Talents Plan.\\

\vskip -3.5mm
\vskip -3.5mm
\section*{APPENDIX A: WAVEGUIDE FABRICATION}
\label{appendix:A}
\vskip -3.5mm
In order to introduce the classical term of the quantum stochastic walk in the seven-waveguide array, we vary $\Delta V$ 40 times in the 8-cm-long evolution length of each waveguide. In other words, $\Delta V$ in each segment of 2~mm is constant, but all of the random $\Delta V$ values in 280 segments (each of the seven waveguides has 40 segments) follow a uniform distribution with a given $\Delta V$ amplitude. All the waveguides are fabricated using the femtosecond laser direct writing technique\cite{Crespi2013,Chaboyer2015,Feng2016}. We steer a 513-nm femtosecond laser (up converted from a pump laser of 10W, 1026~nm, 290~fs pulse duration, 1~ MHz repetition rate) into a spatial light modulator (SLM) to shape the laser pulse in time and spatial domain. We then focus the pulse onto a pure borosilicate substrate with a 50X objective lens (numerical aperture:~0.55). Power and SLM compensation were processed to help ensure the waveguide uniformity\cite{Feng2016}.

In terms of the array arrangment, we originally set it in the 2D planar form, i.e. the sink waveguides and the major waveguides were in the same depth. We found in that case we could not fine tune the characterization as there was always some photonic evolution pattern when the injected light was moved horizontally and even not focused onto the input waveguide, perhaps due to the scattering of the injected light into other waveguides in the same depth. We then set the array in the current 3D form, and we can easily locate the precise evolution facula. This suggests some advantage of the 3D layout.

\vskip -3.5mm
\vskip -3.5mm
\section*{APPENDIX B: $\Delta \beta$ APPROACH}
\label{appendix:B}
\vskip -3.5mm
We measure $\Delta \beta$ using the directional coupler approach. One of the waveguides is written using a base speed $V_0=5~\rm mm/s$, and the other waveguide using a different speed $V$ ($V-V_0=\Delta V$) that will lead to a detuned propagation constant $\Delta \beta$ on the waveguide. In the detuned directional coupler, the effective coupling coefficient $C_{eff}$ can be obtained by using the same coupling mode method as for the normal directional coupler\cite{Szameit2007}, but $C_{eff}$ contains the detuning effect from the $\Delta \beta$ through this equation\cite{Lebugle2015}: $C_{eff}=\sqrt{(\Delta \beta/2)^2+C^2}$, where $C$ is the coupling coefficient for the normal directional coupler. Therefore, $\Delta \beta$ can be calculated when $C_{eff}$ and $C$ are both characterized. We then plot $\Delta \beta$ (unit: $\rm mm^{-1}$) against $\Delta V$ (unit: $\rm mm/s$), and fit it linearly: $\Delta \beta=0.20\times \Delta V$. Knowing this, we can randomly generate $\Delta \beta$ of $0.01-0.4~\rm mm^{-1}$ by varying $\Delta V$ between $0.5-20~\rm mm/s$.

\bigskip

\end{document}